\documentclass[graybox]{svmult}

\usepackage{type1cm}        
\usepackage{makeidx}         
\usepackage{graphicx}        
\usepackage{multicol}        
\usepackage[bottom]{footmisc}

\usepackage{newtxtext}       %
\usepackage[varvw]{newtxmath}       

\makeindex             

\newcommand{\be}{\begin{equation}}
\newcommand{\ee}{\end{equation}}
\newcommand{\beba}{\begin{equation}\begin{array}{lcl}}
\newcommand{\eaee}{\end{array}\end{equation}}
\newcommand{\bea}{\begin{eqnarray}}
\newcommand{\eea}{\end{eqnarray}}
\newcommand{\ba}{\begin{array}}
\newcommand{\ea}{\end{array}}

\def\simlt{\mathrel{\lower2.5pt\vbox{\lineskip=0pt\baselineskip=0pt
           \hbox{$<$}\hbox{$\sim$}}}}
\def\simgt{\mathrel{\lower2.5pt\vbox{\lineskip=0pt\baselineskip=0pt
           \hbox{$>$}\hbox{$\sim$}}}}

\begin{document}

\title*{Extra Dimensions and Physics of Low Scale Strings}
\author{Ignatios Antoniadis  and  Karim Benakli}
\institute{Ignatios Antoniadis \at School of Natural Sciences, Institute for Advanced Study, Princeton, NJ 08540, USA\\
and Sorbonne Universit\'e, CNRS, Laboratoire de Physique Th\'eorique et Hautes Energies, LPTHE, 75005 Paris, France, \email{antoniad@lpthe.jussieu.fr}
\and Karim Benakli \at Sorbonne Universit\'e, CNRS, Laboratoire de Physique Th\'eorique et Hautes Energies, LPTHE, 75005 Paris, France, \email{kbenakli@lpthe.jussieu.fr}}
%
%
\maketitle

\abstract{This review aims to provide a very short and pedestrian introduction to some of the basics of extra-dimensional physics. The hope is to facilitate access and to be, in some respects, complementary to the many already existing reviews on phenomenological applications of extra dimensions in our Universe.}

\section{Introduction}
\label{sec:1}
Is there any compelling reason to believe that there are extra dimensions? If so, why haven't we seen them? Could we discover them in the near future? How would they manifest themselves? These are the questions we face and this review, which is intended to be a very basic introduction to the subject, is never intended to be exhaustive in any sense. We attempt to give an overview of some of the issues studied, which we hope will serve as a starting point for the non-expert reader to explore the vast literature on the subject.

Historically, the possibility of an extra dimension in physics is motivated by the desire to write a theory that unifies gravity with other interactions. This was the idea behind the early work on the subject by Kaluza\cite{Kaluza:1921tu}, Klein \cite{Klein:1926tv,Klein:1926fj}, Einstein and Bergmann \cite{Einstein:1938fk} in particular at the beginning of the 20th century (for a  discussion on some aspects the history of the subject see for example \cite{Witten:2014fma}). Remarkably, their ideas are central, decades later, to the modern approach to the quantification of gravity: string theory. By moving from the notion of fundamental point particles to zero modes of string oscillations, the consistency of our quantum theory then requires us to move to a dimension of space-time greater than four. Indeed, string theory, which offers a framework to unify all interactions, requires the existence of additional degrees of freedom which, in certain limits, take the form of additional dimensions. Let us note here that these dimensions are space and non-temporal in nature, the latter posing problems with the preservation of causality. The possibility of observing extra dimensions has been proposed in several works \cite{Antoniadis:1990ew,Antoniadis:1993jp,Antoniadis:1994yi,Arkani-Hamed:1998jmv,Antoniadis:1998ig}.

Obviously, in a first step, it is natural to start by characterising the additional dimensions by their number: 1, 2, 3, ... their topology, their geometry (flat, curved, wrapped, fractal) and for the physical phenomena we are interested in, the type of states that propagate and interact in them. In general, one is then led to discuss physics in a plethora of spaces of dimensions higher than four, with a huge choice of possible topologies and geometries. We will concentrate here on the simplest configurations that allow us to illustrate some basic examples.

The review is organised as follows: Section 2 presents Kaluza-Klein states. While this is done in the simplest setting of a compactification on a circle, we discuss properties that remain valid in a much larger setting. Section 3 discusses the current experimental limits on the existence of such dimensions. Some applications of these dimensions, relevant for the construction of extensions beyond the Standard Model, are discussed in the rest of the review.

\section{Kaluza-Klein Excitations}
\label{sec:2}

 We need to introduce some basic notions of physics with extra-dimensions useful for our discussions. For this purpose, we shall start by reviewing the simplest model of dimensional reduction from   from  $D+1$ to $D$ dimensions \footnote{We work with the signature $(+,-,...,-)$. The $D+1$ dimensional quantities will be denoted with a hat. We use Latin and Greek letters  for the $D+1$ and $D$-dimensional coordinates, $\hat g$ and $g$  are the determinants of the $D+1$ and $D$-dimensional metric,  respectively. }. We denote the coordinates of the $D$ non-compact  directions by $x^m$ . The $(D+1)-$th dimension is parametrized by  the coordinate $z \equiv z + 2 \pi R$ taken to be a circle of radius $R$. We consider the Einstein-Hilbert action:
\begin{equation}
    \label{action}
    \mathcal{S}_{EH}^{(D+1)}=\frac{1}{2\hat{\kappa}^2}\int\mathrm{d}^{D+1}x \sqrt{(-1)^D\hat{g}}\,\hat{{\cal R}},
\end{equation}
 where $\hat{\cal R}$ is the Ricci scalar. We take for the metric $\hat{g}_{MN}$ the ansatz: 
\begin{equation}
\label{metric}
    \hat{g}_{MN}=\begin{pmatrix}
    e^{2a \phi}g_{\mu \nu}-e^{2 b \phi}A_{\mu}A_{\nu} & e^{2 b \phi}A_{\mu} \\ 
   e^{2 b \phi}A_{\nu} & -e^{2 b \phi}
    \end{pmatrix}
    \end{equation}
    where $\phi$ is the radion/dilaton, $A_{\mu}$ the gravi-photon, $g_{\mu \nu}$ the $D$-dimensional metric and $a,b$ are constants to be determined in the following. It is easy to see that the invariance under general coordinate transformations of $g_{MN}$ becomes gauge invariance of $A_{\mu}$. 

Indeed, the general coordinate transformations of the $(D+1)$-dimensional theory, depending on the vector parameter $\xi^M
$, act on the metric as
\begin{equation}
  \delta_\xi {\hat g}_{MN}= \xi ^P\partial _P {\hat g}_{MN}+ 2 \partial _{(M}\xi ^P {\hat g}_{N)P}\,,
 \label{gctD5}
\end{equation}
where pair of indices in parenthesis are symmetrised with a multiplicative 1/2-factor.
It follows that the field $\phi$ transforms under the $(\xi^\mu =0,\xi^z)$ transformations as
\begin{equation}
  \delta_{\xi^z} \phi  = b\partial_z \xi^z - \xi^z \partial_z \phi \,.
 \label{del4phi}
\end{equation}
This implies that we can gauge-fix the transformations $\xi^z$ for $\partial_z \xi^z\ne 0$ by taking $\phi$ to be independent of $z$:
\begin{equation}
\partial_z\phi=0\,.
 \label{xi4ngauge}
\end{equation}
Similarly, we consider the $\xi^\nu$ transformations of the off-diagonal metric components:
\begin{equation}
  \delta_{\xi^\nu} {\hat g}_{\mu z} = \xi^\nu \partial_\nu {\hat g}_{\mu z} + {\hat g}_{\nu z} \partial_\mu \xi^\nu + {\hat g}_{\mu \nu }\partial_z \xi ^\nu \,. 
\label{delGmu4}
\end{equation}
It follows that we can also gauge-fix the $\xi^\nu$ transformations for $\partial_z \xi^\nu\ne 0$ by taking the off-diagonal components of the metric to be $z$-independent:
\begin{equation}
\partial_z {\hat g}_{\mu z}=0\,.
 \label{ximungauge}
\end{equation}
As a result of the above gauge conditions, in the metric decomposition \eqref{metric}, only the $D$-dimensional metric $g_{\mu\nu}(x^m,z)$ depends on all the $(D+1)$ coordinates, while the radion/dilaton $\phi(x)$ and gravi-photon $A_\mu(x)$ are independent of the compact coordinate $z$.
Moreover, the leftover coordinate transformations depend only of the non-compact coordinates $x^m$: $\xi^\mu(x)$ generate the $D$-dimensional coordinate transformations, while $\xi^z(x)$ acts as a $U(1)$ gauge transformation on the gravi-photon $A_\mu(x)$ that leaves $\phi(x)$ inert. The quadratic in $A$ term in the definition of ${\hat g}_{\mu\nu}$ guarantees that the 0-mode ($z$-independent part) of the $D$-dimensional metric $g_{\mu\nu}$ remains also invariant under $U(1)$ gauge transformations. 

 We now focus on the simplest choice where the $D$-dimensional metric $g_{\mu\nu}$ and thus all fields are independent of the $z$ coordinate. The action \eqref{action} gives then for the zero modes:
   \begin{align}
   \label{actionbeforespecifyingabeta}
 \mathcal{S}_{EH}^{(D+1)}=\frac{2 \pi R}{2\hat{\kappa}^2} \int \mathrm{d}^{D}x\, & \sqrt{(-1)^{D-1}g} \,\, e^{((D-2) a+ b)\phi}\bigg\{{\cal R}-  \big[2(1-D) a -2 \ b \big]\Box \phi 
 \nonumber\\ 
 &\,- \left[ (D-2)(1-D) a^2+2 b \big((2-D) a - b \big)\right](\partial \phi)^2 \nonumber\\ 
 &\,-\frac{1}{4}e^{2( b - a)\phi}F^2\bigg\}.    
   \end{align}
  To have a canonical normalisation of the $D$-dimensional Einstein-Hilbert action, we need 
   \begin{equation}
   \label{relation between alpha and beta}
     (D-2) a+ b=0.  
   \end{equation}
while a canonical dilaton kinetic term requires:
   \begin{equation}
   \label{alpha}
      a^2=\frac{1}{2(D-1)(D-2)}.
   \end{equation}
 In the following we choose the positive root. We can identify the $D$-dimensional Planck mass  
   \begin{equation}
   \label{relation M_P different dimensions}
\frac{1}{\kappa^2}=\frac{2\pi R }{\hat{\kappa}^2} \Longrightarrow  M_P^{D-2}=2\pi R\,\hat M_P^{D-1} \,.
   \end{equation}
  We are used to work with dimensionful fields instead of the dimensionless $\phi$ and $A_{\mu}$. Therefore, we perform the rescaling 
   
   \begin{equation}
   \label{physical fields}
       {\phi} \rightarrow \sqrt{2}\kappa{\phi}{}; \,\,\,\, \,\,\,\, {A_{\mu}}\rightarrow \sqrt{2}\kappa {A_{\mu}}{}
   \end{equation}
leading to the action:
   \begin{equation}
   \label{compactified gravitation action}
       \mathcal{S}^{(D)}\! \!=\! \! \int \! \! \mathrm{d}^Dx\sqrt{(-1)^{D-1}g}\left[ \frac{{\cal R}}{2\kappa^2}+\sqrt{\frac{2}{(D-1)(D-2) }}\Box {\phi}+\frac{1}{2}(\partial {\phi})^2-\frac{1}{4}e^{-2\sqrt\frac{D-1}{D-2}\kappa {\phi}}F^2 \right]
   \end{equation}
where the second term is a total derivative that will not play any role for the purpose of our discussions. 

Let's now consider the corresponding dimensional reduction of a free real massless scalar field $\hat\Phi$:
\begin{equation}
\label{scalar action}
\mathcal{S}_{\Phi}^{(D+1)}=\int\mathrm{d}^{D+1}x\,\,\sqrt{(-1)^{D}\hat{g}}\,\,\frac{1}{2}\hat{g}^{MN}\partial_M \hat\Phi \partial_N \hat\Phi
\end{equation}
Again, for simplicity, we consider the case where the field $\hat\Phi$ is periodic  on the extra dimension
\begin{equation}
\label{scalar decomposition}
    \hat\Phi(x,z+2\pi R)=\hat\Phi(x,z), \qquad \hat\Phi(x,z)=\frac{1}{\sqrt{2\pi R}}\sum_{n=-\infty}^{+\infty}\varphi_n(x)e^{\frac{inz}{R}},
\end{equation}
with $\varphi_{-n}=\varphi^*_n$, which leads to
\begin{align}
\label{action-scalar}
\mathcal{S}_{\Phi}^{(D)}=\int \mathrm{d}^Dx & \sqrt{(-1)^{D-1}g}\Bigg\{\frac{1}{2}\partial_\mu\varphi_0\partial^\mu\varphi_0 +\sum_{n=1}^{\infty}\left(\partial_\mu\varphi_n\partial^\mu\varphi_n^*-\frac{n^2}{R^2}e^{2\sqrt{\frac{D-1}{D-2}}\kappa {\phi}}\varphi_n\varphi_n^*\right) \nonumber \\
&+\sum_{n=1}^\infty \left(i\sqrt{2}\kappa \frac{ n}{R } A^{\mu}\left(\partial_{\mu}\varphi_n \varphi_n^*-\varphi_n\partial_\mu\varphi_n^*\right)+ {2}\kappa^2 \frac{n^2}{R^2}A_\mu A^\mu\varphi_n\varphi_n^* \right)\Bigg\},
\end{align}

We can now comment our results for some generic lessons on extra-dimensions (for a more detailed discussion, see for example \cite{Benakli:2022shq}):

\begin{itemize}
\item The complex scalars $\varphi_n$ form a tower  of massive states with masses
\begin{equation}
\label{mass}
\qquad m_n=\frac{n}{R}e^{\sqrt{\frac{D-1}{D-2}}\kappa {\phi}}.
\end{equation}
This is the Kaluza-Klein (KK) tower. In the very simple case of a circle, the masses are regularly spaced. This is not the case for any compactification. There are however two important generic properties that we would like to note. First the masses depend on the value in vacuum of a scalar field, here the dilaton $\phi$. In the case of supersymmetric theories, these are moduli fields.
Then, in the decompactification limit, by taking $R \rightarrow \infty$ or $\phi \rightarrow -\infty$, there is an infinity of states which become very light exponentially fast. This is one of the manifestations of the Swampland Distance Conjecture in string theory to which we will return below.

\item 
The KK modes appear minimally coupled to the gravi-photon. In general, internal space isometries give rise to gauge symmetries. In the particular case here, it is a $U(1)$ symmetry. The gauge coupling 
\begin{equation}
\label{coupling}
   g^2=e^{2\sqrt{\frac{D-1}{D-2}}\kappa {\phi}}.  
\end{equation}
is again a function of a scalar field, here the dilaton $\phi$, and taking $g \rightarrow 0$, thus $\phi \rightarrow -\infty$, we get an infinity of light states, again manifestations of the Swampland Distance Conjecture.

 The charges of the KK states are given by
\begin{equation}
\label{charge and mass}
    gq_n=\sqrt{2}\kappa \frac{n}{R}e^{\sqrt{\frac{D-1}{D-2}}\kappa {\phi}} 
\end{equation}
Mass and charge are related through
\begin{equation}
\label{relation mass charge}
    (gq_n)^2=2\kappa^2m_n^2,
\end{equation}
saturating the BPS condition. This is expected as all the interactions unify to descend from the unique gravitational interaction of a free scalar field in higher dimensions.

\item 
The same tower of KK modes appears also in the massive spin-2 excitations of the metric $g_{\mu\nu}(x^\rho,z)$ in the matrix decomposition \eqref{metric}. Note that in the unitary gauge choice where the radion/dilaton $\phi(x)$ and the gravi-photon $A_\mu(x)$ are 
independent of the internal compact coordinate $z$, only the metric $g_{\mu\nu}$ has KK excitations. The would be $(D-2)+1$ excitations of $A_\mu$ and $\phi$ are absorbed by the KK modes of the metric to provide the longitudinal and scalar helicities of the massive spin-2. Indeed, the latter has $(D-1)D/2$ helicities, versus $(D-2)(D-1)/2$ of the massless case.

\item Finally, we have established in the equation a relation between the Planck masses in $D$ and $D+1$ dimensions. This relation is generalized in the case of the factorization of any internal space with the Minkowski spacetime. For a $\delta$-dimensional internal space $K$ with volume $Vol(K)$, we have:
\begin{equation}
\label{relation mass charge 2}
 \frac{1}{\kappa^2}=\frac{Vol(K)}{\hat{\kappa}^2} \Longrightarrow  M_P^{D-2}=Vol(K)\,\hat M_P^{D+\delta-2}      
 \end{equation}
which for the case where $K$ is a torus of equal radii $R$ reads:
 \begin{equation}
\label{Torus mass relations}
\frac{1}{\kappa^2}=\frac{(2\pi R)^\delta}{\hat{\kappa}^2} \Longrightarrow  M_P^{D-2}=(2\pi R)^\delta\,\hat M_P^{D+\delta-2}      
 \end{equation}
This  implies in particular that the large value of the $D$ dimensional Planck mass (compared with a given energy scale, for example the electroweak scale) might be a consequence of large volume while the $D+\delta$ dimensional Planck mass is much smaller. 
\end{itemize}

This last observation has important consequences. For example, as the four-dimensional Planck scale $ M_P$ is related to the "fundamental Planck scale" $\hat M_P$ through
\begin{equation}
\label{Torus mass relations 2}
M_P^{2} \sim  R^\delta\,\hat M_P^{2+\delta} \,,     
 \end{equation}
the existence of Large Extra Dimensions with size $R {\hat M_P} \gg 1$ allows for a low fundamental scale.  The fundamental scale can be as low as the TeV \cite{Arkani-Hamed:1998jmv,Antoniadis:1998ig}, or at intermediate energies \cite{Benakli:1998pw,Burgess:1998px,Montero:2022prj}.

 Before mentionning experimental investigations for extra dimensions, we still need to discuss two important aspects of compactifications.

One aspect is called $T$-duality. When the size of a dimension tends to zero, the KK modes become very massive. However, in string theories, there are states whose mass decreases when the size of additional dimensions decreases. These are the winding states, whose masses are given by:
\begin{equation}
\label{winding masses}
{\tilde m_l}=l R e^{-\sqrt{\frac{D-1}{D-2}}\kappa {\phi}}\,,
\end{equation}
where the integer $l$ is the winding number of the closed string around the compactification circle. The simplest form of $T$-duality corresponds to the inversion of the compactification radius (in string units) together with the exchange of the KK and winding modes
. It turns out that this is a symmetry of the closed string spectrum and interactions.

The other aspect concerns the existence of points in the inner space where certain states are localised. These can be particular points in this space, but can also be edges. From the point of view of point particle theory, these points seem to correspond to singularities, but these are resolved in string theory. Examples  are either fixed points of an orbifold, or branes. The simplest example of such a case is an internal space described as a segment.  In this case, some states would be localised at its edges. An important property then is that since these points break the translation invariance in the corresponding directions of the internal space, the momentum along them is not conserved. Therefore single KK modes can decay to, or be produced from, states localised there. For completeness, we describe below the main properties of orbifold compactifications, restricting to the simplest case of a line interval.

{\it Orbifolds:} The main motivation comes from generating chirality, since toroidal compactifications give rise to non-chiral theories since a five dimensional spinor is reduced to a Dirac fermion in four dimensions. Emergence of chirality requires internal manifolds with non-trivial holonomy, while orbifold are special (singular) limits obtained by quotient of a regular manifold by a discrete symmetry. The resulting spaces have conical singularities at the fixed points of the discrete symmetry transformations. The simplest one-dimensional example is the line interval $I$, that can be obtained by quotient a circle $S^1$ by the $Z_2$ parity $z\to -z$, $I=S^1/Z_2$. The two end-points of the interval, $z=0$ and $z=\pi R$, are fixed points under $Z_2$.

Since $Z_2$ is a symmetry of the higher dimensional theory, five dimensional fields should be even or odd under the $Z_2$ parity. They can thus be expanded in terms of $\cos nz/R$ or $\sin nz/R$, instead of plane waves in \eqref{scalar decomposition}:
\begin{align}
\label{even-odd decomposition}
\hat\Phi_{\rm even}(x,-z)=\hat\Phi_{\rm even}(x,z) &:\quad 
\hat\Phi_{\rm even}(x,z)=\frac{1}{\sqrt{\pi R}}\varphi_0^e(x) + \sqrt{\frac{2}{\pi R}}\sum_{n=1}^{+\infty}\varphi_n^e(x)\cos{\frac{nz}{R}}
\nonumber\\
\hat\Phi_{\rm odd}(x,-z)=-\hat\Phi_{\rm odd}(x,z) &:\quad \hat\Phi_{\rm odd}(x,z)=\sqrt{\frac{2}{\pi R}}\sum_{n=1}^{+\infty}\varphi_n^o(x)\sin{\frac{nz}{R}}\,,
\end{align}
where the normalisation is fixed for $z\in [0,\pi R]$.
Note that only the even fields have four-dimensional zero-modes. Actually, there is an alternative way of studying orbifolds by imposing boundary conditions at the `end of the world' boundaries, which in our example are the two end-points of the interval. 
The even fields correspond to Neumann boundary conditions where the derivative of the fields vanish at the end-points, while the odd fields correspond to Dirichlet boundary conditions where the fields vanish at the end-points:
\begin{align}
\label{boundary conditions}
\partial_z\hat\Phi_{\rm even}(x, z=0) &=\partial_z\hat\Phi_{\rm even}(x, z=\pi R)=0\\
\hat\Phi_{\rm odd}(x, z=0) &=\hat\Phi_{\rm odd}(x, z=\pi R)=0
\end{align}

Consider next a 5-dimensional gauge field ${\hat A}_M(x,z)$ associated with a gauge transformation: 
\begin{equation}
\label{gauge field}
\delta_\omega {\hat A}_M(x,z)=\partial_M\omega(x,z)\,. 
\end{equation}
Choosing the gauge parameter $\omega$ to be even under $z\to -z$, it follows that ${\hat A}_\mu$ is even while ${\hat A}_z$ is odd. Moreover, we can gauge-fix the $\omega$ transformation for $\partial_z\omega\ne 0$ by taking $\partial_z A_z=0$, implying that $A_z$ vanishes since it is odd. As a result, one is leftover with ${\hat A}_\mu$ which has a KK mode expansion as an even field in \eqref{even-odd decomposition} with its 4-dimensional vector 0-mode $A_\mu(x)^{(0)}$ associated to a 4-dimensional gauge transformation $\omega(x)$.

Let us consider now a 5-dimensional massless fermion $\hat\Psi(x,z)$; its transformation under $Z_2$ has the form:
\begin{equation}
\label{fermions}
\hat\Psi(x,-z)=\Gamma\,\hat\Psi(x,z)\quad;\quad \Gamma^2=1\,.
\end{equation}
Since the vector current $\overline{\hat\Psi}\gamma^M\hat\Psi$ must transform as ${\hat A}_M$, $\overline{\hat\Psi}\gamma^\mu\hat\Psi$ must be event while $\overline{\hat\Psi}\gamma^5\hat\Psi$ must be odd. It follows that $\Gamma=\gamma^5$, implying that the 4-dimensional zero-mode of  $\hat\Psi$ is chiral. 
Thus, orbifold compactifications lead to chirality.




\section{Searches for Extra Dimensions}
\label{sec:3}

In the previous section we have pointed out manifestations of extra-dimensions: the KK modes and modification of gravitational interaction by the presence of a lower fundamental Planck mass. We shall use these two facts here to search for experimental signatures of extra dimensions.

\subsection{Table-Top Experiments}
\label{subsec:1}
From now on, we consider a spacetime of dimension $4+d$ which factorizes as 3+1 non-compact dimensions times an internal space $K$ of dimension $d$. 

At large length scales, the typical energies of particles are far too small to produce on-shell Kaluza-Klein modes. No signal propagates in $K$, the space-time appears four-dimensional. By increasing the energy we eventually reach values sufficient to create KK excitations and we then start to probe the $K$ space. At length scales much
smaller than the radii of  $ \delta$ compact dimensions, among the $d$ of $K$,  the gravitational interactions are locally well approximated by a $4+ \delta$ dimensional description. Applying the Gauss law, one expects a transition from a gravitational potential from an $1/r$ behavior to an $1/r^{1+\delta}$. At lengths comparable to the compactification scale, the gravitational potential can be parametrized as:
\begin{equation}
V(r)=- G_N \frac{m_1 m_2}{r}(1 + {\alpha} e^{-r/{\lambda}})\,,
\end{equation}
where $\lambda=1/m$ is the wavelength of the lighter KK mode of mass $m$.
For $ \delta$ extra dimensions, one has:
\begin{equation}
{\alpha}= \frac{8 \, \delta}{3}\,,
\end{equation}
providing the strength of the extra Yukawa-type force relative to gravity.

Many tabletop experiments have been designed to investigate the existence of new forces at very short distances (for a review see for example \cite{Adelberger:2003zx}). In particular, the tightest limits for extra forces of strength comparable to gravity $(\alpha\sim{\cal O}(1))$ are obtained by torsional pendulum kind of experiments. Basically, in this type of experiments, the idea is that a torsion pendulum is holding a disc placed over another rotating attractor disc. The upper disc is put in movement by twisting the cable of the pendulum, while the lower one is rotating at constant speed. Due to the presence of holes on the discs, there is a contribution to the torque when all the holes are not facing each other. The hanging disc feels a periodic torque that can be measured with a very high precision. In fact,  one measures a set of oscillation frequencies and compare with expectation. The measured torques are now consistent with a purely inverse square law interaction and thus allow us to put bounds on new forces and possible large extra-dimensions.

The forces we are trying to measure accurately are gravitational in nature and therefore very weak. There are many sources of background and noise to contend with, which one needs to be able to evaluate theoretically and reduce their importance in the experiments. We do not discuss them in detail here. Instead, we just very briefly mention how the treatment of certain background sources limits the domain that can be probed by these experiments. Firstly, in order to avoid the presence of Coulomb forces that cannot be evaluated, conductive materials are used. Secondly, to minimise the effects of potential differences between the crystals at different locations on the facing surfaces, a gold coating is used. Finally, to reduce all these effects as well as the effects of Casimir forces, the detector and attractor surfaces are moving relative to each other, but also a stiff, stationary, conductive membrane is placed between them.  To improve the minimum separation between the pendulum and the attractor has been a challenge because of  sensitivity to alignment uncertainties, vibrations and even dust particles.The thickness of the membrane means however that it is not possible to go to distances of less than about ten microns. All together with an important improvement in modeling and data analysis during the years has allowed to get to an impressive sensitivity. The strongest limit for $ \delta =2$ extra dimensions, is today of order $R \lesssim 30 \mu m$ \cite{Kapner:2006si,Tan:2020vpf}.

\subsection{Astrophysics and Cosmology Bounds}
\label{subsec:2}

Models with extra dimensions are subject to cosmological and astrophysical constraints \cite{Arkani-Hamed:1998sfv,Benakli:1998ur,Hall:1999mk}. 

The first constraint is that the gravitons in the bulk should not store too much energy. Relic gravitons should not close the Universe: This constrains for example the fundamental scale to $\simgt 7 $ TeV for $ \delta =2$ extra-dimensions. 

Then these KK relic gravitons can decay into photons and contribute to the cosmic diffuse gamma ray. By imposing that the models do not predict a too bright sky, one obtains limits on the fundamental scale, e.g. for $ \delta =2$ extra dimensions the fundamental scale of the theory must be $\simgt 100$ TeV \cite{Benakli:1998ur,Hall:1999mk}.

Fairly light KK excitations of gravitons are abundantly produced in the interior of stars. These particles carry a quantity of energy which can be non-negligible. The agreement of stellar models with observations leads to limits on the number of such KK states. The limit on the fundamental scale from the supernova SN1987A is of order of 27 TeV for $ \delta =2$ extra dimensions \cite{Hanhart:2001fx}. After a supernova explosion, the least energetic KK gravitons remain gravitationally trapped in the remaining neutron star. The requirement that neutron stars should not be excessively heated by the decays of these KK modes into photons gives the strongest limit such as 1740 TeV for $ \delta =2$ extra dimensions~\cite{Hannestad:2003yd}, excluding this scenario as a solution for the electroweak versus Planck scales hierarchy problem.

\subsection{Searches at Collider Experiments}

Analyses of the data collected from experiments at colliders allow to put other bounds on the size of extra dimensions \footnote{ A detailed discussion can be found in the Particle Data Group section on extra dimensions \cite{ParticleDataGroup:2022pth}.}. The main searches can be separated into two categories: searches for  Standard Model particles, in particular gauge bosons \cite{Antoniadis:1994yi,Antoniadis:1999bq}, their KK excitations on one side and and graviton KK modes \cite{Giudice:1998ck,Mirabelli:1998rt,Han:1998sg,Hewett:1998sn}  on the other side. In each case, one can search either for on-shell production or for virtual exchange effects. With increasing energy, the string substructure can also be excited, leading to stringy effects at colliders. These can be on-shell production, or give rise to dimension 6 or 8 effective operators.

The graviton KK excitations produced in colliders interact very weakly and thus appear as missing transverse energy. Following the non-observation of this missing energy in jets production, ATLAS \cite{ATLAS:2021kxv} and CMS \cite{CMS:2021far} give limits of the same order, with 139 (137) fb$^{-1}$
\begin{equation}
M_{P,4+\delta} \gtrsim 5.5 - 11   \, {\rm TeV} \qquad \delta =6 - 2.
\end{equation}

ATLAS \cite{ATLAS:2019fgd} and CMS \cite{CMS:2019gwf} bounds on the fundamental scale, from non-observation of string resonance modes in di-jets, are:
\begin{equation}
M_{string} \gtrsim 8  \, {\rm TeV}
\end{equation}

A priori, one might think that a limit on the fundamental scale can be derived by considering that the exchange of virtual modes of KK gravitons will induce an effective operator of the form:
\begin{equation}
\frac{4}{\Lambda}(T_{\mu \nu}T^{\mu \nu}- \frac{1}{\delta +2}T_{\mu}^{\mu}T^{\nu}_{\nu})
\end{equation}
where $T_{\mu \nu}$ is the stress-energy tensor and $\Lambda$ is of the order of the fundamental Planck mass. CMS (2018) bounds the size of this operator from analysis of the dijet angular distribution with 35.9 fb$^{-1}$ to be of order \cite{CMS:2018ucw}:
\begin{equation}
\Lambda \gtrsim 9.1  \, {\rm TeV} 
\end{equation}
 However, this limit should not be taken too seriously. In any quantum theory of gravity, one expects to see new states, new interactions at this scale.  These will induce new effective operators of dimension 8, or worse of dimension 6, which will pollute or dominate the virtual graviton exchange operators.  This has been studied in particular for string theory in \cite{Accomando:1999sj,Cullen:2000ef,Antoniadis:2000jv}. These operators are therefore too polluted to extract robust bounds.
 
 Another prediction of extra-dimensional models is the existence of KK excited states of Standard Model particles. We will not discuss here in detail the different possibilities, but we will comment on the case of KK gauge bosons. 

The simplest case corresponds to the KK modes which appear as a sequential tower of Z', W', gluons' and photons' which interact 
with localized states at the boundary. Limits were derived from the analysis of the 2018 LHC data (36 fb-1). Rescaling the ATLAS \cite{ATLAS:2019lsy} and CMS \cite{CMS:2018hff} experiments bounds would give a limit roughly in the range $\frac{1}{R} \gtrsim 10 \, {\rm TeV}$.

 \subsection{   How many KK modes?}
    
    In the case of on-shell production of KK modes, their number is determined by the available energy and phase space. But, in the case of an exchange of virtual KK modes, one has an infinite number that can be exchanged. At first sight, the contribution of the sum then seems to diverge \cite{Antoniadis:1993jp}. Indeed, let us consider the sum over the propagators, for example of modes exchanged in the $s \rightarrow 0$ channel of a $2 \rightarrow 2$ interaction. This reads:
\begin{equation}
\sum_{n_1, \cdots, n_d} \frac{g_{\vec n}^2}{-s+m_0^2+ {n_1^2+ \cdots +n_d^2}}
\end{equation}
 This sum diverges for  $d>1$ if the couplings $g_{\vec n}$ of the KK modes are independent of ${\vec n}$. Often in the literature, one finds that the sum is made finite by cutting-off the number of KK states, to take into account only those below the fundamental Planck mass. However, as we will see in the next section, the whole infinite tower of KK modes is required in order to preserve all symmetries of the theory and such a truncation is not justified. We then turn to string theory where such sums exist and give finite and consistent results. Let us suppose for instance that we exchange the KK excitations of gauge bosons~\cite{Antoniadis:1993jp}:
\begin{equation}
A^\mu (x,\vec y)=\sum_{{\vec n}}
{\cal A}^{\mu }_{\vec n}(x) \exp{i\frac {n_i y_i}{R_i}}
\end{equation}

The result can be described by an effective theory where the coupling of the gauge bosons to the localized charge current takes the form \cite{Antoniadis:2000jv}:
\begin{equation}
\int d^4x \, \, \,  \, \sum_{{\vec
n}} e^{-\ln {\Delta} \sum_i\frac{n_i^2l_s^2}{2 R_i^2}} \, \,
\, \, \, j_\mu (x) \, {\cal A}^{\mu }_{\vec n}(x)\, ,
\end{equation}

After Fourier transform, this reads:
\begin{equation}
\int d^{4}y \,\int d^4x  \, \,
\, \,  \left(\frac{1}{l_s^2 2 \pi \ln {\Delta}}\right)^{2} e^{- \frac {{\vec
y}^2}{2 l_s^2 \ln {\Delta}}}  \, j_\mu (x) \, A^\mu (x,\vec y)\,
\end{equation}

The localized matter is then felt as a Gaussian distribution of charge:
\begin{equation}
e^{-\frac {{\vec y}^2}{2 \sigma^2}}
 j_\mu (x) \qquad \sigma=\sqrt{\ln {\Delta}}\, l_s \sim 1.66 \, l_s
 \end{equation}
where $\Delta>1$ is a model/compactification dependent parameter defining the localisation width. A similar parameter enters in the coupling of KK gravitons to D-branes.




\section{Symmetry Breaking Through Extra-Dimensions}
\label{sec:4}

Suppose that to hide the extra dimensions one "chooses" \footnote{ We have seen above that the size of the extra dimensions is a vacuum expectation value  of the redion field, which must be dynamically determined by the minimization of a corresponding scalar potential.} a very small size for the compactification radii. The natural scale would of course be of the order of the string scale $\frac{1}{R} \sim \, M_{s}$, which is then roughly $\frac{1}{R} \sim \, M_{P}$. The isometries of the compactification space allow us to obtain gauge symmetries, for example, in Section~\ref{sec:2} we obtained a $U(1)$ gauge symmetry. Unfortunately, as noted by Oskar Klein, a quantized elementary charge implies particles of mass $\sim \frac{1}{R} \sim M_{P}$. Note, however, that this is a consequence of having imposed periodicity of the fields along the internal directions, an unnecessary condition as we will see for two cases in this section.

\subsection{Wilson Lines and Gauge Symmetry Breaking}
\label{subsec:4}

For a scalar field $\Phi$, the four-dimensional effective potential 
can be written in  the Schwinger representation as:
\bea
V_{eff}(\Phi)&=&\frac{1}{2} \sum_I (-)^{F_I} \int \frac{d^4p}{(2\pi)^4}
\log\left[p^2+M_I^2(\Phi)\right] 
\cr
&=&-\frac{1}{2} \sum_I (-)^{F_I} \int_0^\infty 
\frac{dt}{t}   \int\frac{d^4p}{(2\pi)^4} \,  e^{-t\left[ p^2+M_I^2(\Phi)
\right]}\cr
&=&-\frac{1}{32 \pi^2} \sum_I (-)^{F_I} \int_0^\infty 
\frac{dt}{t^3} \, \, e^{-t M_I^2(\Phi)}
\cr
&=&-\frac{1}{32 \pi^2} \sum_I (-)^{F_I} \int_0^\infty 
dl \, \, l   \, \, e^{- { M_I^2(\Phi)}/{l}}
\label{potential}
\eea
where the sum is over all bosonic ($F_I = 0$) and fermionic ($F_I = 1$) 
degrees of freedom with $\Phi$-dependent masses $M_I(\Phi)$ and we have made the change of variables $t=1/l$. It is useful to keep in mind that the 
ultraviolet (UV) and infrared (IR) contributions correspond to the integration
regions $t \rightarrow 0$ ($l\rightarrow \infty$) and 
$t \rightarrow \infty$  ($l \rightarrow 0$), respectively.

Here we are interested in a peculiar case of scalar fields: Wilson lines that descend from the dimensional reduction of a vector $\hat A_{M}$ in higher dimensions\footnote{ The material presented here follows the work \cite{Antoniadis:2001cv,Benakli:2007zz}.} . To illustrate this, let us consider the action:
\begin{equation}
\mathcal{S}=\int\mathrm{d}^{5}x\,\,\sqrt{(-1)^D\hat{g}}\,\, \left\{ \hat D_M \hat\Phi \hat D^M \hat\Phi^*-\hat M^2\hat\Phi\hat\Phi^* -\frac{1}{4}\hat F_{M\,N}\hat F^{M\,N}\right\},
\end{equation}
where  $\hat F_{M\,N}$ is the field strength of $\hat A_{M}$. We consider that  $\hat A_{M}$ is periodic 
\begin{eqnarray}
\hat A_M(x,z+2\pi R)=\hat A_M(x,z) \Rightarrow & \hat A_M(x,z)=\frac{1}{\sqrt{2\pi R}}\sum_{n=-\infty}^{+\infty}A_{(n) M}(x) e^{\frac{inz}{R}}  \end{eqnarray}
while $\Phi$ changes by a phase:
\begin{eqnarray}
\label{z-dependence}
    \hat \Phi(x,z\!+\! 2\pi R)\!=\! e^{2i\pi q_\Phi\omega}\hat \Phi(x,z) \Rightarrow & \hat \Phi (x,z)=\frac{1}{\sqrt{2\pi R}}\sum_{n=-\infty}^{+\infty} \varphi_n(x)e^{i(n+q_\Phi\omega)\frac{z}{R}},
\end{eqnarray}
where $q_\Phi$ is the $U(1)$ charge of $\Phi$ and $\omega$ an arbitrary constant. This boundary condition is allowed since the 5-dimensional theory has a global $U(1)$ symmetry (as part of the gauge symmetry).
Let us now consider a non-vanishing internal component of the gauge field $ <A_{(0) 5}(x)> = h_0 \neq 0$. The mass term in the Lagrangian for the scalar KK modes takes then the form:
\begin{equation}
\left(\hat M^2+\left[\frac{n+a}{R}\right]^2\right)\left|\varphi_n\right|^2
\label{Wilson-Mass} 
\end{equation}
where $a$ is Wilson line given by:
\begin{equation}
\label{adef}
a= q_\Phi \omega -q_\Phi g h_0 R=q_\Phi \left(\omega - g\oint \frac{dy^i}{2 \pi} A_i\right)\,.
\end{equation}

It is straightforward to generalise to the case of $d$ extra dimensions compactified on
 circles with radii $R_i >1$, in units of the fundamental (string) length scale, 
with $i=1, \dots ,d$. Then, equation \eqref{Wilson-Mass} becomes
\bea
M^2_{\vec{m},I}= M^2_{I}(\phi)+ \sum_{i=1}^{d}
\left[ \frac {m_i+a^{I}_{i}(\phi) } {R_i}\right]^2
\label{mastermass}
\eea
where $I$ labels bosonic, but also fermionic, fields  and  ${\vec{m}} =\{m_1,\cdots ,m_d \} $ with $m_i$  integers, while the $y^i$ coordinates parametrise the $d$-dimensional torus.

In the following, we take $M^2_{I} =0$, leading to the one-loop effective potential:
\bea
V_{eff}(\phi)|_{\rm torus} =- \sum_I \sum_{\vec{m}} (-)^{F_I} 
\frac{1}{32 \pi^2}  \int_0^\infty 
dl \, \, l \, \, e^{- \sum_i \frac { \left(m_i +a^{I}_{i} 
\right)^2}{ R_i^2 \, l }}\,.
\label{before}
\eea
We start by commuting the integral with the sum over the KK states. As the number of KK states is infinite, we can perform a Poisson resummation:
\bea
V_{eff}(\phi)|_{\rm torus}=-\sum_I (-)^{F_I} \, 
 \frac{\prod_{i=1}^{d} R_i}{32\,  \pi^{\frac{4-d}{2}}}
\sum_{\vec{n}}  e^{ 2 \pi i \sum_i n_i a^{I}_{i}}\, \,   \int_0^\infty 
dl \, \,  l^{\frac{2+d}{2}} 
\,  \,  e^{- \pi^2 l \sum_i n_i^2 R_i^2}
\label{after}
\eea
The $\vec{n} = \vec{0}$ contribution is a cosmological constant which is divergent in this simple case, but irrelevant for the purpose of this review.
For all $\vec{n} \neq \vec{0}$, 
 the change of variables: $l'= \pi^2\, l\, \sum_i n_i^2 R_i^2$ and 
integration over $l'$  leads to:
\begin{equation}
V_{eff}(\phi)|_{\rm torus} = -\sum_I (-)^{F_I} \, \, 
\frac{\Gamma(\frac{4+d}{2})}{32 \pi^{\frac{12+d}{2}}}\,  \,
\prod_{i=1}^{d} R_i  \,  \,
 \sum_{\vec{n}\neq \vec{0}}  \frac {e^{ 2 \pi i \sum_i n_i a^I_i(\phi)}}  
{\left[ \sum_i n_i^2 R_i^2 \right]^{\frac{4+d}{2}}}
\label{finaltor}
\end{equation}
which is {\it finite}   for the $\phi$-dependent part of the 
effective potential and, thus, computable in the field theory limit.
It is important to stress again the importance, illustrated by this computation, to keep the whole infinite tower of KK modes and not to truncate them at some UV cut-off of the effective field theory.

The gauge field ${\hat A}_M$ could in general be part of a non-abelian group. In this case, Wilson lines can be turned on along the Cartan generators leading to breaking patterns of the gauge group associated to vacuum expectation values in the adjoint representation.

\subsection{Coordinate Dependent Compactification and Supersymmetry Breaking}
\label{subsec:5}

It may happen that in particular compactifications, the internal scalar component of the higher-dimensional gauge field does not survive as physical excitation in the spectrum, but still discrete values are allowed associated to a discrete symmetry of the theory. In this case, $h_0$ in \eqref{adef} is absent and the parameter $\omega$ takes discrete values. Still, this leads to gauge group symmetry breaking, even in the absence of the corresponding Higgs scalars in the spectrum. The net effect can also be interpreted as discrete Wilson lines. This symmetry breaking by boundary conditions is also called `coordinate dependent' (or Scherk-Schwarz) compactification~\cite{Scherk:1978ta,Scherk:1979zr}, since zero modes of charged fields acquire $z$-internal coordinate dependance, as it can be seen in the KK-mode expansion \eqref{z-dependence}.

The above mechanism of gauge symmetry breaking can be generalised to any symmetry, such as supersymmetry that we discuss here. The relevant symmetry one can use to break supersymmetry is R-symmetry that acts as a phase rotation of the fermionic coordinates in the superspace. As a result, bosons and fermions which are different superfield components transform differently and have different boundary conditions \eqref{z-dependence}, breaking supersymmetry. In general, such as in string theory, global continuous R-symmetry is broken by the compactification to a discrete subgroup, so that the supersymmetry breaking scale is not a free parameter, independent from the compactification, or the string scale~\cite{Antoniadis:1988jn}.

The simplest and universal example is a $Z_2$ parity that corresponds to the fermion number. Thus, fermions change sign and, thus, $q\omega=1/2$ in \eqref{z-dependence} leading to KK-modes with half-integer frequencies, corresponding to $a=1/2$ in \eqref{Wilson-Mass}. On the other hand, bosons are invariant and their KK-modes remain unaffected. The resulting supersymmetry breaking spectrum is identical to the Matsubara frequencies at finite temperature, where the compactified space coordinate $z$ is replaced by time with the radius being replaced by the inverse temperature. Bosons are periodic with integer frequencies, while fermions are antiperiodic leading to half-integer frequencies~\cite{Rohm:1983aq,Kounnas:1988ye,Antoniadis:1990ew}.

It turns out that coordinate dependent compactifications are exemples of freely acting orbifolds without fixed points, implying the absence of boundary states (associated to twisted sectors). This is illustrated in a simple $Z_2$ example described below.

{\it Freely acting orbifold:} Let us replace the interval $S^1/Z_2$ by combining the parity transformation $z\to -z$ with the Kaluza-Klein parity $(-)^n$. This amounts to shifting $z$ by $\pi R$ and eliminates the fixed points since the end-points of the interval are now exchanged by the orbifold action. Unlike the KK-mode decomposition \eqref{even-odd decomposition} in the $S^1/Z_2$ orrdifold, now both even and odd fields under $z$-parity can have cosine and sine expansions, depending on the KK-parity which is even for $n=2k$ even integers and odd for $n=2k-1$ odd integers:
\begin{align}
\label{freely acting KK}
\hat\Phi_{\rm even}(x,z)=&\frac{1}{\sqrt{\pi R}}\varphi_0^{e,e}(x)\\
& + \sqrt{\frac{2}{\pi R}}\sum_{k=1}^{+\infty}\varphi_{2k}^{e,e}(x)\cos{\frac{2kz}{R}}
+\sqrt{\frac{2}{\pi R}}\sum_{k=1}^{+\infty}\varphi_{2k-1}^{e,o}(x)\sin{\frac{(2k-1)z}{R}}\nonumber\\
\hat\Phi_{\rm odd}(x,z)=& \sqrt{\frac{2}{\pi R}}\sum_{k=1}^{+\infty}\varphi_{2k-1}^{o,e}(x)\cos{\frac{(2k-1)z}{R}}
+\sqrt{\frac{2}{\pi R}}\sum_{k=1}^{+\infty}\varphi_{2k}^{o,o}(x)\sin{\frac{2kz}{R}}\,,\nonumber
\end{align}
where the second superscript of the mode functions $\varphi_n$ refer to the even $(e)$ or odd $(o)$ transformation with respect to the KK-parity that is combined with the action of the $z$-parity of the 5-dimensional field and $z$-parity of the wave functions (even for cosine and odd for sin) to form an overall even action under $Z_2$.

{\it Scherk-Schwarz  compactification:} Let us consider now a Scherk-Schwarz compactification described previously based on a $Z_2$ symmetry acting on the 5-dimensional fields and imposing boundary conditions as in \eqref{z-dependence} with $q\omega=1/2$, so that odd fields are antiperiodic. This leads to the following KK-mode expancions:
\begin{align}
\label{SS-KK}
\hat \Phi_P (x,z)&=\frac{1}{\sqrt{2\pi R}}\sum_{n=-\infty}^{+\infty} \varphi^P_n(x)e^{in\frac{z}{R}}\\ 
&=\frac{1}{\sqrt{2\pi R}}\varphi_0^{P}(x) +\frac{1}{\sqrt{\pi R}}\sum_{k=1}^{+\infty}\varphi_{k}^{P,R}(x)\cos{\frac{2kz}{2R}}
+\frac{1}{\sqrt{\pi R}}\sum_{k=1}^{+\infty}\varphi_{k}^{P,I}(x)\sin{\frac{2kz}{2R}}
\nonumber\\
\hat \Phi_A (x,z)&=\frac{1}{\sqrt{2\pi R}}\sum_{n=-\infty}^{+\infty} \varphi^A_n(x)e^{i(n+\frac{1}{2})\frac{z}{R}}
\nonumber\\
&= \frac{1}{\sqrt{\pi R}}\sum_{k=1}^{+\infty}\varphi_{k}^{A,R}(x)\cos{\frac{(2k-1)z}{2R}}
+\frac{1}{\sqrt{\pi R}}\sum_{k=1}^{+\infty}\varphi_{k}^{A,I}(x)\sin{\frac{(2k-1)z}{2R}}\,,\nonumber
\end{align}
where $P$ $(A)$ refers to periodic (antiperiodic) fields, while the superscripts $R$ and $I$ stand for the real and imaginary part of the corresponding mode function.

It is now easy to see that the expressions \eqref{freely acting KK} and \eqref{SS-KK} can be identified upon doubling the radius from the interval in \eqref{freely acting KK} to the circle in \eqref{SS-KK} and the identification:
\begin{align}
\label{identification}
\varphi^P_k=\varphi^{e,e}_{2k}-i\varphi^{o,o}_{2k}\quad;\quad \varphi^A_k=\varphi^{o,e}_{2k}-i\varphi^{e,o}_{2k}\,.
\end{align}

\section{Extra-Dimensions and the Swampland Conjectures}
\label{sec:5}

Additional dimensions play a central role in string theory. Indeed, one does not have to live in the flat space-time of the critical string dimensions, but by means of a different `dimensional reduction', one can restrict the number of infinite dimensions of the space where the effective field theory lives. Several properties of the known string theory compactification models have recently been promoted as properties of all quantum theories of gravity by a series of conjectures. Two conjectures are of particular interest here. The first is the Distance Conjecture stating that in directions of large distances in the moduli space, there is a tower of light states with masses exponentially small with the proper distance measured in Planck units. The second is the Weak Gravity Conjecture stating that gravity is the weakest force. This implies the existence of a state with charge bigger than its mass in Planck units, allowing black holes to decay, evading stable remnants.

\subsection{Very weak couplings}

The Swampland Conjectures \cite{Vafa:2005ui,Arkani-Hamed:2006emk} forbid arbitrarily small gauge coupling. This is if we can take it to zero, we get a global symmetry which is argued to be absent in quantum gravity.  We will illustrate the obstruction to taking this limit in an example, that of the dark photon.

In string theory, one can suppress the strength of gauge coupling $g_X$  by taking  $\delta_X$ extra dimensions to be large. Indeed, one can express $g_X$   as

\begin{equation}
g_X^2 =\frac{(2 \pi)^{\delta_X +1} \ g_s}{V_X \ M_s^{\delta_X}} 
\end{equation}
or equivalently as   
  
\begin{equation}
g_X^2 = 2 \pi g_s \left( \frac{8}{g_s^{2}} \right)^{\delta_X/d}  \ \left(\frac{M_s}{M_{\rm Pl}} \right)^{2 \delta_X/d}  
\end{equation}
 
Taking all the compact dimensions large ($\delta_X =d$), we get \cite{Benakli:2020vng,Anchordoqui:2020tlp} 
\begin{equation}
g_X = \sqrt{\frac{16 \ \pi}{g_s}} \ \frac{M_s}{M_{\rm Pl}}
\sim  4 \times10^{-14}   \, \, {\left( \frac{0.2}{g_s} \right)^{1/2}} \left(\frac{M_s}{10 \, \, {\rm  TeV}} \right)\,
\end{equation}
For $g \sim 10^{-14}$, we see that new physics would appear at the scale: $\Lambda_{UV} \lesssim M_s\sim g M_{\rm Pl} \sim 10 \, {\rm TeV}$. This is the order of magnitude of the smallest string scale allowed by experiments, as we have seen above. So much small values of gauge couplings are not allowed in such theories. Moreover, independently of experiments, we see that the limit $g_X \rightarrow 0$ implies $\Lambda_{UV}  \rightarrow 0$, and there is no more energy region where the effective theory is valid. This is an illustration of Magnetic Weak Gravity Conjecture.

On the other hand, taking $\delta_X =0$, or alternatively chosing $V_X$ of string size $\sim M_S^{-\delta_X}$, one obtains $g_{X} \sim \sqrt{g_s}$ while $M_s\sim g_s M_{\rm Pl}$ when all compactification sizes are of order the string length. Thus, $g_{X}$ can because small by choosing a tiny string coupling $g_s$. A realisation of this possibility is within the limit of little string theory~\cite{Antoniadis:2001sw,Antoniadis:2011qw}, or using heterotic small instantons~\cite{Benakli:1999yc}.

Consider heterotic strings compactified on a $K3$ fibered over a two-dimensional base $P^1$ such that the approximation 
  \begin{equation}
 <V_{K3} V_{P^1}> \simeq <V_{K3}>< V_{P^1}>
  \end{equation}
is valid. Here, $V_{P^1}$  and  $V_{K3}$ denote the  volume of $P^1$ and $K3$, respectively.
If all the compact dimensions are large: 
\begin{equation}
M_{\rm Pl}^2 = \frac{64 \ \pi}{g_s^2} \ M_s^8 \ <V_{K3} V_{P^1}>
\end{equation}
We assume here that the Standard Model states arise from small instantons localised on $K3$. The volume $V_{P^1}$ is then forced to remain of order one to avoid suppressing the Standard Model gauge couplings\cite{Benakli:1999yc}. We have \cite{Anchordoqui:2020tlp} 
  \begin{equation}
g_{X}^2 = \frac{g_s}{2} \frac{ 1}{M_s  <V_{P^1}>} =  4
\sqrt{\pi}  \frac{M_s}{ M_{\rm Pl}} M_s^2\sqrt{<V_{K3}>} \sim 6 \times 10^{-14} \frac{M_s}{  10\ {\rm TeV}} 
  \end{equation}
where in the last estimate we used a string size $K3$, $ <V_{K3}>\sim M_s^{-4}$.
We can get now tiny gauge coupling for the dark photon not by large $V_{K3} $ but also taking a tiny $g_s$. For instance: $g_s \sim 10^{-13}$ leads to $M_s \sim 10 \, {\rm TeV}$.

We conclude that tiny couplings request either very large extra dimensions or a low string scale, therefore an infinite tower  of KK states or of string resonances. The weaker the coupling, the lower the energy at which these towers appear.

\subsection{ The Dark Dimension}
 
We very briefly mention the recent proposal for one extra dimension of length in the micrometer range. There are indications that if one continuously takes the limit of an AdS space  cosmological constant towards zero, there is a tower of states that become very light with masses that go parametrically like $ |\Lambda |^{-1/4}$~\cite{Lust:2019zwm}.  In~\cite{Montero:2022prj}, the authors assume that dS solutions will behave in this respect as AdS spaces. Further, they assume that these tower of states have the description of KK modes of an extra dimension.  Given the tiny size of our Universe dark energy, these states are very light   and do not decouple from the EFT:

\begin{equation}
m \sim \lambda' \, \,  \Lambda^{\frac{1}{4}}, \qquad \lambda' \sim 10 - 10^3  
\end{equation} 

This requires $R \sim 0.1 - 10 \, \mu m  $ and corresponds to a fundamental scale of order $\sim 10^9 - 10^{10} \, {\rm GeV} $ which falls in the intermediate energies region mentioned above.

%
%
%

\end{document}